\documentstyle[12pt]{article}

\def\d{\delta} 
\def\l{\lambda} 
\def\o{\omega} 
\def\t{\tau} 
\def\bt{{\bar \tau}} 
\def\vh{v_r^H(z)}

\def\o{\omega}
\def\z{\zeta}
\def\bz{{\bar \zeta}}

\def\cn{{\cal N}}

\newcommand{\be}{\begin{equation}}
\newcommand{\ee}{\end{equation}}
\newcommand{\bea}{\begin{eqnarray}}
\newcommand{\eea}{\end{eqnarray}}

\def\nonu{\nonumber \\{}}

\begin{document}


\newcommand{\nd}[1]{/\hspace{-0.5em} #1}

\begin{titlepage}

\begin{flushright}

{\bf
May 2002} \\ 
SWAT-339  \\ 
hep-th/0205151 \\

\end{flushright}

\begin{centering}

\vspace{.2in}

 {\large{\bf ${\cal N}=1^{*}$ vacua, Fuzzy Spheres and Integrable Systems
}}\\

\vspace{.4in}

 Nick Dorey and Annamaria Sinkovics  \\

\vspace{.4in}

Department of Physics, University of Wales Swansea \\
Singleton Park, Swansea, SA2 8PP, UK\\

\vspace{.2in}



%

%

\vspace{.4in}

{\bf Abstract} \\

\end{centering}

We calculate the exact eigenvalues of the adjoint scalar fields 
in the massive vacua of ${\cal N}=1^{*}$ SUSY Yang-Mills with gauge
group $SU(N)$. This provides a field theory prediction for the
distribution of D3 brane charge in the AdS dual. We verify the
proposal of Polchinski and Strassler that the D3-brane's lie on a
fuzzy sphere in the supergravity limit and determine the corrections
to this distribution due to worldsheet and quantum effects. The
calculation also provides several new results concerning the
equilibrium configurations of the $N$-body Calogero-Moser 
Hamiltonian.



\end{titlepage}

\section{Introduction}

The ${\cal N}=1^{*}$ deformation of ${\cal N}=4$ SUSY Yang-Mills
theory with gauge group $SU(N)$ 
provides a fascinating example of a non-conformal gauge theory
where a direct comparison of the AdS dual with field theory results is
possible. Specifically, as this is a theory with ${\cal N}=1$ SUSY, 
there are special observables which depend holomorphically on the
parameters and can be computed exactly for all values of the marginal
coupling $\tau=4\pi i/g^{2}+\theta/2\pi$ and $N$. 
In the limit of large 't Hooft coupling, 
$\lambda=g^{2}N/4\pi \rightarrow \infty$, the results can be compared with 
a supergravity computation in the IIB dual proposed by Polchinski and
Strassler \cite{ps}. 
\paragraph{}

The ${\cal N}=1^{*}$ theory contains three complex 
scalar fields $\Phi_{1}$,
$\Phi_{2}$ and $\Phi_{3}$, in the adjoint representation of
$G=SU(N)$. After a rescaling, these fields appear on an equal
footing and may be rotated into each other by an $SO(3)$ subgroup of
the $SU(4)$ R-symmetry of the ${\cal N}=4$ theory. 
In the following we will study the eigenvalues of these 
$N \times N$ matrices in supersymmetric vacuum states of the theory,
or equivalently, the vacuum expectation values of the $N-1$ 
gauge-invariant operators $u_{k}={\rm Tr}_{N}\, \Phi^{k}$ for
$k=2,\ldots, N$. Here, and in the following, $\Phi$ denotes any one of
the three $\Phi_{i}$ and $\lambda_{k}$, with $k=0,1,\ldots,N-1$, denotes
its eigenvalues. The main result of this paper is an exact formula for
the large-$N$ distribution of these eigenvalues in each massive vacuum
state of the theory. As we explain below, the desired result 
may be obtained by considering the equilibrium configurations of a 
certain classical integrable system. 
In the remainder of this introductory section we
will give this result and discuss its physical
interpretation. Details of the calculations are provided in subsequent
sections. 
\paragraph{}

We begin by reviewing the classical vacuum structure of the theory. 
The classical F-term vacuum equations are
solved when the three adjoint scalars (appropriately rescaled) 
obey an $SU(2)$ algebra
$[\Phi_{i},\Phi_{j}]=\varepsilon_{ijk} \Phi_{k}$. In fact, as first
explained in \cite{VW}, the classical theory has one SUSY vacuum for each 
$N$-dimensional representation of $SU(2)$, with the VEVs of the 
$\Phi_{i}$ being given by the anti-Hermitian 
generators acting in this representation. 
An arbitrary representation of dimension $N$ can be decomposed as the 
sum of irreducible representations whose dimensions add up to
$N$. The total
number of classical vacua is therefore equal to the number of
partitions of $N$. 
The unique irreducible representation corresponds to a vacuum where
the gauge group is completely broken by the Higgs mechanism and the
spectrum is massive. We will
refer to this as the `Higgs vacuum'. In vacua corresponding to
reducible representations, a non-trivial subgroup of $SU(N)$ is 
left unbroken and there are massless particles at the classical
level. 
\paragraph{}

This correspondence between classical ${\cal N}=1^{*}$ vacua and $SU(2)$
representations has an interesting realization in weakly coupled IIB
string theory where the undeformed ${\cal N}=4$ theory with gauge group 
$U(N)$ lives on the
worldvolume of $N$ parallel D3 branes. The ${\cal N}=1^{*}$
deformation corresponds to turning on the flux of the two-form
antisymmetric tensor field of the Ramond-Ramond sector of the IIB
theory. The background RR flux causes the D3 branes to polarize to 
spherically
wrapped D5 branes by the Myers effect \cite{myers}. As there is non-zero 
D3-brane
charge on the two wrapped dimensions of the fivebrane, the corresponding
worldvolume theory is non-commutative. In field theory, the
non-commutivity is realized by the $SU(2)$ algebra described above 
satisfied by the VEVs of the adjoint scalar fields, $\Phi_{i}$. 
The $N$-dimensional irreducible representation corresponding to the 
Higgs vacuum is the familiar 
matrix representation of the fuzzy sphere. Reducible representations
correspond to multiple concentric fuzzy spheres in an obvious way. 

\paragraph{}
With the standard normalization of the generators, the eigenvalues of 
each of
the adjoint scalars in the Higgs vacuum is uniformly distributed
on the imaginary axis between $\pm i (N-1)/2$ 
with unit spacing. Explicitly, the $N$ eigenvalues are
$\lambda_{k}=i(N-2k-1)/2$ with $k=0,1, N-1$. In Section 2.2, we will
derive an exact quantum mechanical formula for the 
eigenvalues $\lambda_{k}$ in the Higgs vacuum (see equation
(\ref{hv1}). In the large-$N$ limit we
replace $k/N$ by a continuous variable $0\leq x<1$. The range of the
eigenvalues determines the radius of the sphere on which the D3 branes
lie. Equivalently, it is also determined by the expectation value of the lowest
non-zero condensate $u_{2}$ which has already been studied in detail in
earlier work \cite{nick,np,ADK}. For the present purposes 
it is convenient to normalize the range of the eigenvalues to unity, 
defining a
normalized eigenvalue distribution $\tilde{\lambda}(x)$. (The natural
normalization as we later show involves the dual 't Hooft coupling
and the modular weight.) The large-$N$
limit of the semiclassical result therefore yields a uniform
distribution $\tilde{\lambda}(x) =1-x/2$. In the following we will see
that this uniform distribution also emerges in the strong-coupling
supergravity regime.  
\paragraph{}

So far we have only described the classical limit of the ${\cal
N}=1^{*}$ theory. In vacua with an unbroken non-abelian gauge
symmetry, the low-energy theory is realized in a strongly-coupled 
confining phase with gluino condensation and a mass gap. The 
vacuum structure of the quantum theory is determined by the exact
superpotential derived in \cite{nick}. 
\begin{equation}
{\cal W}= m_{1}m_{2}m_{3}\sum_{a>b} {\cal P} (X_{a}-X_{b})
\label{esp}
\end{equation}
Here, ${\cal P}(X)$ is the Weierstrass function and $m_{i}$, $i=1,2,3$, 
are the masses of the three chiral multiplets which we will set to 
one from now on. The $N$
auxiliary variables $X_{a}$, with $a=1,2,\ldots,N$, only have a direct
physical interpretation when the theory is compactified on a circle
down to three dimensions. In this context they can be identified with
a complex combination of the Wilson and 't Hooft loops of the gauge
field around the compact direction. These variables naturally take
values on a torus whose complex structure is the complexified coupling
constant $\tau$, and the exact superpotential,
being an elliptic function, is single-valued on this torus. The
S-duality transformations of the ${\cal N}=4$ theory correspond to
modular transformations of the torus. 
\paragraph{}

Stationarizing the superpotential 
with respect to the variables $N$ variables $X_{a}$, 
we find a set of massive vacua labeled
by integers $p$,$q$ and $k$, with $pq=N$ and $k=0,\ldots, q-1$. 
The total number of these vacua is the sum of the divisors of $N$. 
For each divisor $q$ of $N$, the $q$ quantum vacua labelled by
$k=0,\dots, q-1$  reduce in the weak coupling limit to the 
unique classical vacuum where the unbroken gauge symmetry is $SU(q)$. 
This classical vacuum corresponds to the reducible representation of
$SU(2)$ which is the direct sum of $q$ copies of the irreducible
representation of dimension $p=N/q$. The Higgs vacuum 
corresponds to $q=1$, $p=N$. 
For $q>1$, the low energy theory
is strongly coupled and gluino condensation occurs, splitting the
massless classical vacuum into $q$ massive quantum vacua. 
The theory also has a large
number of vacua which remain massless at the quantum level but we will
not consider these here. 
\paragraph{}

A remarkable feature of the ${\cal N}=1^{*}$ theory is that the set of
$\sum_{d|N} d$ vacua described above are transformed into each other 
by the S-duality of the underlying ${\cal N}=4$ theory. In particular,
the electromagnetic duality transformation $\tau\rightarrow -1/\tau$ 
interchanges the Higgs vacuum in which electric charges condense 
and a confining vacuum in which magnetic charges condense. This
provides a concrete realization of the duality between
Higgs and confining phases conjectured many years ago by 't Hooft. 
This property is manifest in the exact results following from 
the superpotential (\ref{esp}). The action of S-duality can also be
understood in the IIB realization of the theory described
above. As above the Higgs vacuum is realized on $N$ parallel D3 branes
polarized into a spherically wrapped D5-brane. The D3 branes are
invariant under the $SL(2,Z)$ duality of IIB string theory, but the D5
brane is not. The action of S-duality generates an $SL(2,Z)$ orbit of
vacuum configurations described by $(m,n)$-fivebranes wrapped on
non-commutative two-spheres each carrying $N$ units of D3-brane
charge. Here $m$ and $n$ are determined in terms of the the integers $p$, $q$
and $k$ which describe the field theory vacua \cite{ps}. 

\paragraph{}
Polchinski and Strassler studied the AdS duals of these 
${\cal N}=1^{*}$ vacua. They proposed a dual IIB geometry 
for each vacuum which is asymptotically $AdS_{5}\times S^{5}$ near
the boundary, but is deformed by the presence of the spherically
wrapped $(m,n)$-fivebranes in the interior. The precise domain in
which the supergravity approximation is reliable depends explicitly on
the choice of vacuum. We start from the vacuum which exhibits pure
magnetic confinement, corresponding to the choice $p=1$,
$q=N$ and $k=0$. This vacuum is described by a single spherically wrapped NS
five-brane. The SUGRA description is valid when the sphere is large in
string units and this is true as long as the `electric' 't Hooft 
coupling, $\lambda=g^{2}N/4\pi$, is much larger than one. Away from this
limit corrections from IIB worldsheet instantons wrapping the $S^{2}$
are unsuppressed. Part of the charm of this model is that these
corrections can be calculated explicitly in the dual field theory.   
As explained above the remaining vacua are related to the magnetic
confinement vacuum by S-duality. Specifically, we perform an
$SL(2,Z)$ transformation, $\tau\rightarrow
\tau_{D}=(a\tau + b)/(c \tau +d)=4\pi
i/g_{D}^{2}+\theta_{D}/2\pi$ where $a$, $b$, $c$, $d$ are integers with
$ad-bc=1$ which transforms the wrapped NS five-brane into an $(a,c)$ 
five-brane. In this vacuum we define a dual 't Hooft coupling 
$\lambda_{D}=g_{D}^{2}N/4\pi$ and the classical supergravity approximation is
reliable whenever $\lambda_{D}>>1$ with $g^{2}_{D}<<1$. Away from
large $\lambda_{D}$, corrections coming from $(a,c)$-strings wrapping
the two-sphere become significant. 
\paragraph{}
In the 
regime described above, the supergravity dual can be used to
compute various field theory quantities such as the tensions of BPS
domain walls and the VEVs of chiral operators. In 
\cite{ps,ADK}, these were successfully
compared with the exact field theory results of
\cite{nick,np}. However, the
particular quantities tested so far correspond to operators of low
$R$-charge such as the gluino condensate and the operator $u_{2}$
defined above. These operators correspond to massless scalars
present
in five-dimensional gauged supergravity. 
One of the main aims of this
paper is to extend this test to operators of higher R-charge
corresponding to Kaluza-Klein modes which probe the full ten-dimensional
geometry. Specifically, the eigenvalues $\lambda_{k}$ determine
the VEVs of the operators $u_{k}$, which have charge proportional to
$k$ under a $U(1)$ subgroup of the $SU(4)$ R-symmetry of the ${\cal
N}=4$ theory. The corresponding SUGRA states live in the tower   
of spherical harmonics on the five-sphere of the UV geometry. 
\paragraph{}

The VEVs of the operators $u_{k}$ can be read off from the asymptotic 
behaviour of these SUGRA fields in the Polchinski-Strassler
solution corresponding to each vacuum. 
Indeed, this was the proceedure followed in \cite{ADK} to compute the
VEV of $u_{2}$. However it is simpler to note that, as at
weak-coupling, the eigenvalues $\lambda_{k}$ determine the position of
the distribution of D3 brane charge in spacetime. The correspondece
between these two points of view for the ${\cal N}=4$ theory 
is discussed in
 
\cite{krauss}. By calculating the $\lambda_{k}$ directly in field
theory and then taking the SUGRA limit, we will be able to check
directly that the 3-branes lie on non-commutative two-spheres in
spacetime as predicted by Polchinski and Strassler. Our main result is
a universal formula for the large-$N$ eigenvalue distribution which
is valid in all the massive vacua

\be
\tilde{\l}(x) =  \frac{1}{2} - x  +
\sum_{l=1}^{\infty}
\left( \frac{e^{- 2 \pi \l_D (l - x )}}{ 1 +
e^{- 2 \pi \l_D (l - x)}}  -
\frac{e^{-2 \pi \l_D (l - 1 + x )}}{ 1 +
e^{- 2 \pi \l_D (l - 1 + x )}} \right) \label{fform1}
\ee
Here, as above $\lambda_{D}=g^{2}_{D}N/4\pi$ is the appropriate 't
Hooft coupling in each vacuum. In the limit $\lambda_{D}\rightarrow
\infty$ only the first term on the right-hand side contributes and we
find a uniform distribution of eigenvalues centered at the origin
$x=0$. This is consistent with
the fuzzy sphere structure described above. In particular, we recall
that the eigenvalues of the generators of $SU(2)$ in an irreducible
representation are uniformly distributed around the origin. The
corrections correspond to the contributions of wrapped $(a,c)$-strings
around this sphere \cite{ps,ADK}. 
\paragraph{}

Our approach to determining the exact eigenvalues $\lambda_{k}$ is based on the connection
between supersymmetric gauge theories and integrable systems first
discovered in \cite{DW,R,MW}.  
In particular, it is noteworthy that the 
exact superpotential coincides with the 
potential energy of the elliptic Calogero-Moser system. The latter is
an integrable model describing non-relativistic particles moving
two-body interaction potential determined by the Weierstrass
potential. In terms of the positions and momenta of the $N$ particles,
$X_{a}$ and $P_{a}$ with $a=1,\ldots, N$, the Hamiltonian is, 
\begin{equation}
H=\sum_{a=1}^{N}\, \frac{P_{a}^{2}}{2} \,+ \, 
\sum_{a>b} {\cal P}\left(X_{a}-X_{b}\right)  
\label{ham}
\end{equation}
Although the positions and momenta are real in the first instance, it
is convenient to promote them to complex variables. 
\paragraph{}
In its original form the connection between integrable systems and 
supersymmetric gauge theory involves theories with ${\cal N}=2$ 
supersymmetry. The theory relevant to the present case is the ${\cal N}=2$ 
supersymmetric deformation of the ${\cal N}=4$ theory, also known as 
the ${\cal N}=2^{*}$ theory. In particular
the complex curve which governs the Coulomb branch of this theory
coincides with the {\it spectral curve} of the Calogero-Moser system
which plays a key role in its explicit integration. As explained in
\cite{nick}, the fact that the
exact superpotential of the ${\cal N}=1^{*}$ theory coincides with the
Calogero-Moser Hamiltonian follows from soft breaking to ${\cal N}=1$ 
supersymmetry. In particular, the supersymmetric vacua are in one to
one correspondence with the equilibrium positions of the (complexified)
Calogero-Moser system. The results given below extend the
existing results in the literature about the equilibrium properties of
the integrable system. In particular, they provide an elliptic
extension of the results given in \cite{cal} for the trigonometric
Calogero-Sutherland model.  
\paragraph{}

For the present purposes, certain aspects of the correspondence
between the ${\cal N}=1^{*}$ system and the Calogero-Moser system are 
particularly useful. The integrability of the Calogero-Moser system is
equivalent to the existence of $N$ conserved quantities or
Hamiltonians, $H_{k}$ with $k=1,2,\ldots N$, 
of which the first two are the total momentum and
energy. As discussed in the next section, Hamilton's equations 
of motion for
the system (\ref{ham}) have a Lax Representation and the $N$ conserved
quantities are determined by the eigenvalues of the {\it Lax matrix}
$L$ as $H_{k}=Tr_{N} L^{k}$. On the other hand, the identification of the
Calogero-Moser spectral curve with the complex curve of the ${\cal
N}=2^{*}$ theory reveals that the Hamiltonians $H_{k}$ should be
identified with the VEVs of the operators\footnote{The vanishing of
$u_{1}$ in the $SU(N)$ theory corresponds to choosing the center of
momentum frame $H_{1}=\sum_{a=1}^{N}\, p_{a}=0$ in the integrable
model.} $u_{k}={\rm Tr}_{N} \Phi^{k}$. This gives us a very convenient
way to calculate the eigenvalues $\lambda_{k}$ in each SUSY
vacuum: they can be identified with the eigenvalues of the Lax matrix in the
corresponding equilibrium position of the Calogero-Moser system. 
\paragraph{}

The identification described above is complicated by the problem of 
operator mixing as discussed in \cite{ADK}. In particular, once 
masses are turned on, the operator 
$u_{j}$ can mix with the other operators $u_{k}$ with $k<j$ which have 
lower dimensions. This is particularly clear in the ${\cal N}=2^{*}$ theory, 
where the ambiguity is simply that of choosing holomorphic coordinates on 
the Coulomb branch. Similarly, in the integrable system, it is possible 
to find many inequivalent Lax matrices whose eigenvalues correspond to 
different linear combinations of the conserved quantities.  Hence there is 
no unique way to the operators $u_{k}$ in terms of the conserved 
Hamiltonians. Fortunately this amibiguity turns out to be very mild in the 
large-$N$ limit.  As explained in \cite{ADK} the ambiguity is vacuum 
independent and can be fixed by determining the mixing coefficients in 
any given vacuum. These are holomorphic functions of the 
complexified coupling which necessarily have a weak coupling expansion in 
integer powers of the Yang-Mills instanton factor 
$q_{YM} =\exp 2\pi i \tau$. 
These terms vanish exponentially in the 't Hooft limit, and the 
mixing coefficients are pure numbers which can be determined by a 
semiclassical calculation. The upshot is that once we have determined the 
semiclassical relation between the eigenvalues of the 
Lax matrix and the field theory operators in the Higgs vacuum we can make 
definite statements about the large-$N$ distribution of D3 branes 
in the other vacua.       

\paragraph{}
The paper is organised as follows. In Section 2 we introduce the
integrable system and determine the eigenvalues of the Lax matrix in
the equilibrium configurations. In Section 3, we check the prediction
of Polchinski and Strassler that the eigenvalues lie on a fuzzy sphere
in each vacuum. Section 4 we perform a detailed analysis of the
results in the confining vacuum and in Section 5 we compare with
earlier results for the condensate $u_{2}$. Another approach to 
calculating the field theory VEVs, based on the complex curve of the 
${\cal N}=2^{*}$ theory is described in an appendix.

\section{The Calogero-Moser integrable system}

The CM system \cite{op} consists of $N$ non-relativistic particles
with position $X_a$ 
which interact 
via the two-body potential
\be
V= \sum_{a>b} {\cal P}(X_a - X_b) \quad a,b =1 \ldots N
\ee
where ${\cal P}$ is the Weierstrass function defined on a torus with 
periods $2\omega_{1}=2i\pi$ and $2\omega_{2}=2i\pi \tau$. 
The coordinates and momenta are considered complexified and the 
positions of particles are centered so that 
\be
\sum_{a=1}^N X_a =0
\ee
The system has a non-trivial equilibrium position at,
\be
X_a = \frac{ 2 i \pi a}{N}  -  \frac{i \pi (N-1)}{N} \quad a=1 \ldots N
\ee
where the $N$ particles are equally spaced along the one-cycle of the 
torus parallel to the period $2\omega_{1}$.  
As explained in \cite{nick}, this configuration corresponds to the 
Higgs vacuum of the ${\cal N}=1^{*}$ theory. Acting by modular 
transformations, we find a family of equilibrium configurations equal in 
number to the sum of the divisors of $N$ less one where the
particles are equally space along different cycles of the torus.
These correspond to field theory vacua in the confining phase.

The Calogero-Moser system can be described in terms of a pair of 
Lax matrices, and is
characterized by $N$ integrals of motion and corresponding 
angular variables. The $\cn=1^*$ vacua are described in terms of the 
$N$ order parameters in the moduli space $u_{k}={\rm Tr}\Phi^k$, 
which correspond to 
the $N$ integrals of motion of the integrable system modulo the potential 
ambiguities described above.
The integrals of motions or Hamiltonians for the system
are the symmetric polynomials of the eigenvalues of the Lax matrix $L$.

\paragraph{}
Hamilton's equations for system of $N$ particles are equivalent to the
Lax matrix equation \cite{op}

\be
i \dot{L} = [M, L]
\ee
where 
\bea
L_{ab} &=& p_a \d_{ab} + i (1 - \d_{ab}) x(X_a -X_b) \nonu
M_{ab} &=& \d_{ab} \sum_{a \ne c} z(X_a -X_c) - (1 - \d_{ab}) y (X_a - X_b)
\eea
with 
\bea
V(u) &=& x(u)^2 + C \nonu
y(u) &=& - x'(u) \nonu
z(u) &=& \frac{x''(u)}{2 x(u)} \label{xlax}
\eea
where $C$ is a constant. 
In the present case where the potential $V$ is the Weierstrass 
function, $x(u)$ is given in terms of Jacobi 
elliptic functions \cite{op}. There is more then one possible 
choice for the function $x(u)$. 
Different choices lead to different possible Lax matrices whose eigenvalues 
lead to distinct linear combinations of the $N$ conserved quantities.    
A particularly convenient choice is, 
\be
x(u) = \frac{{\bf K}}{i \pi} cs \left({\bf K} u / {i \pi}  \right)
\ee
where $cs(u)$ is defined as the ratio of the Jacobian elliptic functions 
$cs(u) = cn(u) / sn(u)$,  and ${\bf K}$ and 
${\bf K'}$ are complete elliptic integrals.   
The square of $x(u)$ is periodic in $u$ with periods 
\bea
2 \o_1 &=& 2 \pi i \nonu
2 \o_2 &=& - \frac{2 \pi {\bf K}'}{{\bf K}} = 2 \pi i \tau
\eea
so 
\be
\frac{{\bf K}'}{{\bf K}} = -i \tau
\ee

The Lax matrix in the Higgs vacuum equilibrium position has the form 
\be
L_{ab} = ( 1 - \delta_{ab}) 
\frac{{\bf K}}{ \pi} cs \left( {\bf K} u_{ab} / {i \pi} \right)  
\label{laxgen}
\ee
where 
\be
u_{ab} = \frac{2 \pi i}{N} (a-b)
\ee
Conveniently, the matrix L is a circulant, and 
its eigenvalues are computed by a standard formula as, 
\bea
\l_a &=& \sum_{m=1}^{N-1} A_m z_a^m \nonu
A_m &=&-\frac{{\bf K}}{\pi} cs \left( 
 2 {\bf K} m / N \right)
\nonu
z_a &=& e^{2 \pi i a / N}
\eea
Using the peridic properties of $cs(u)$,  
the eigenvalues can be rewritten in the form of a discrete Fourier transform
\be
\l_a = i \sum_{m=1}^{N-1} A_m \sin{ \frac{2 \pi a m}{N} } 
\ee
In the continous large $N$ limit this becomes a continuous Fourier transform,
and the eigenvalues are the Fourier coefficients of the function 
$cs(u)$. Here we note that the Lax matrix is manifestly traceless.

\subsection{Semiclassical limit}

Before giving a general formula for the eigenvalues in the Higgs vacuum, 
we examine
the semiclassical limit of the integrable system, 
$g \rightarrow 0$ limit $\t \rightarrow i \infty$.  
In this limit potential becomes trigonometric,   
\be
V = \sum_{a>b} \left( \frac{1}{4 \sinh^2{ \frac{1}{2} (X_a - X_b) }} +
\frac{1}{12} \right)
\ee
The equilibrium configuration for the system coincides with the
configuration corresponding to the Higgs vacuum in the
general case. The Lax matrix for this system \cite{op} in the equilibrium 
position reduces to, 
\be
L_{ab} = \frac{1}{2} ( 1 - \delta_{ab})
\cot{ \frac{\pi (a -b)}{N}  } \label{lax1}
\ee 
with the Lax matrix function $x(u)$
\be
x(u) = \frac{1}{2} \coth{\frac{u}{2}}
\ee
We have again chosen the Lax matrix function $x(u)$ for which 
$L$ is a circulant, and so is convenient for the computation for the 
eigenvalues. Comparing the form of the Lax matrix to the general case 
(\ref{laxgen}), the general CM system is recognized as the elliptic 
deformation of the integrable system with $1/\sinh^2$ potential 
(Calogero-Sutherland system).

The eigenvalues are computed as
\be
\l_a = \frac{1}{2} \sum_{m=1}^{N-1} A_m z_a^m
\ee
where
\bea
A_m &=& -\cot{ \frac{\pi m}{N} } \nonu
z_a &=& e^{2 \pi i a / N}
\eea
$\l_a$ can also be rewritten as
\bea
\l_a &=& -\frac{i}{2} \sum_{m=1}^{N-1} \cot{ \frac{\pi m}{N}} \sin{ \frac{ 2 \pi a m}{N} }
= -\frac{i}{2} (N - 2 a)  \quad a = 1 \ldots N-1  \nonu
\l_{N} &=& 0 \label{evalsem}
\eea
The summation identity used here is the first one in a series of 
diophantine identitities first given in \cite{cal}. 
In general these relations are highly non-trivial and conderations of 
special properties of the Lax matrix of the Calogero-Sutherland system 
actually provided the proof. In this particular case the result is not hard to prove by elementary means. 
\footnote{An easy way to prove the relation is considering the
difference of two consecutive eigenvalues.} In the following we will derive 
elliptic deformations of these identities.

The above formula for the eigenvalues of the Lax matrix, nearly matches the 
classical field theory formula $\lambda_{k}=-i(N-2k-1)/2$. In semiclassical 
field theory, the $N$ eigenvalues of $\Phi$ are evenly distributed 
along the imaginary axis with unit spacing centered around the origin. 
In (\ref{evalsem}), only $N-1$ eigenvalues are distributed in this 
way and there is an extra zero eigenvalue. This discrepancy is 
suppressed by a power of $1/N$ and the two eigenvalue distributions 
agree in the large-$N$ limit. This supports our identification of the 
eigenvalues of the Lax matrix with the field theory VEVs. As explained 
in the introduction, this identification can now be used to yield 
predictions for the D3 brane distributions in the other vacua which are 
unambiguous in the large-$N$ limit.

\subsection{General formula for the eigenvalues}

A general formula for the eigenvalues arises using the Fourier expansion
of the function $cs(u)$, 
\be
cs(u)
= \frac{\pi}{2 {\bf K}} \left[ \cot{\frac{\pi u}{2 {\bf K}}}
- 4 \sum_{n=1}^{\infty} \frac{q^{2 n}}{1 + q^{2n}} 
\sin{\frac{n \pi u}{{\bf K}}}
\right]
\ee
where
\be
q = e^{- \pi \frac{{\rm K}'}{{\rm K}}} = e^{i \pi \t}
\ee
is the factor associated with a single Yang-Mills instanton. 

Substituting the expansion 
\be
\l_a = -\frac{i}{2} \sum_{m=1}^{N-1} \cot{\frac{\pi m}{N}} 
\sin{ \frac{2 \pi a m}{N} } + 2 \sum_{m=1}^{N-1} \sum_{n=1}^{\infty} 
\frac{q^{2 n}}{1 + q^{2n}} \sin{\frac{2 \pi n m}{N}}
e^{2 \pi i a m /N}
\ee

The first piece in the summation gives back the semiclassical result
for the eigenvalues (\ref{evalsem}). The remaining terms are instanton
corrections. 
\paragraph{}
Next we evaluate the eigenvalues for the general case keeping the
full series in the summation. Since the sum is absolutely convergent, 
the order of the summation is interchangable. 
The double summation then can be performed by rewriting
\be
\l_a 
= -\frac{i}{2} (N - 2 a) - i \sum_{n=1}^{\infty} \sum_{m=1}^{N-1}
\frac{q^{2 n}}{1 + q^{2n}} \left( e^{ \frac{2 \pi i m}{N} (a+n)} -
e^{ \frac{2 \pi i m}{N} (a-n)} \right)
\ee
Summing the resulting exponentials, the first summation selects
\bea
n  &=& l N  -a \quad l \ge 1 \nonu
n &=&  a - l N  \quad  l \le 0
\eea 
and the sum becomes
\bea
\l_a &=& -\frac{i}{2} (N - 2 a) - i N \left( \sum_{l=1}^{\infty} 
\frac{q^{2 (l N  -a)}}{1 + q^{2(l N  -a)}}  - \sum_{l=-\infty}^{l=0} 
\frac{q^{2 (a - l N )}}{1 + q^{2(a - lN )}} \right)  \nonu
&=& -\frac{i}{2} (N - 2 a) - i N \sum_{l=1}^{\infty} \left( \frac{q^{(2l-1)N} 
q^{N-2 a}}{ 1 + q^{(2l-1)N} q^{N-2 a}} - 
\frac{q^{(2l-1)N} q^{2 a -N}}{1 + q^{(2l-1)N} q^{2 a -N}} \right)
\label{hv1}
\eea
The summation is recognized as the expansion of the derivative of logarithmic 
Jacobi theta function
\bea
\frac{\theta_3'(z|q)}{\theta_3(z|q)} =  2 i \sum_{l=1}^{\infty} \left(
\frac{q^{2 l -1} e^{2 i z}}{1+ q^{2l-1} e^{2 i z}} - 
\frac{q^{2 l -1} e^{-2 i z}}{1+ q^{2l-1} e^{-2 i z}} \right)
\eea
The final result of the summation can be written as
\be
\l_a = -\frac{i}{2} (N - 2 a) - \frac{1}{2} N
\frac{\theta_3'(z_a|\tilde{q})}{\theta_3(z_a|\tilde{q})} \label{res}
\ee
with
\bea
\tilde{q} &=& q^N = e^{N i \pi \t} \nonu
z_a &=& \frac{1}{2} (N- 2 a) \pi \t
\eea
This is a general formula for the eigenvalues as a function of $N$ and
the effective coupling in the Higgs vacuum $\tilde{\tau}=N\tau$.

\section{Confining vacuum}

In this section we compute the eigenvalues in the unique vacuum of the
theory which exhibits purely magnetic confinement.

As reviewed in Section 1, the confining vacuum in question is 
the image of the Higgs vacuum considered in the previous section under
the electromagnetic duality transformation 
$\t \rightarrow - \frac{1}{\t}$. As the moduli $u_{k}$ each have
holomorphic modular weight $k$, the individual eigenvalues should
transform with modular weight one. 

Under the S transformation the Jacobi theta function transforms
according to Jacobi's formula
\be
\theta_3(z|\t) = (- i \t)^{-\frac{1}{2}} e^{z^2 / \pi i \t}
\theta_3 \left( \left. \frac{z}{\t} \right| -\frac{1}{\t} \right)
\ee
Performing the S transformation including the modular weight 
we get for the eigenvalues in the confining vacuum
\be
\l_a = \frac{i}{2 \t} (N - 2 a) + 
\frac{N}{2 \t} \frac{\theta_3'(\tilde{z}_a|\tilde{q})}{\theta_3(\tilde{z}_a|\tilde{q})}
\label{confo} \ee
where $\tilde{\t} = \frac{\t}{N}$ is the effective coupling in the
confining vacuum, in terms of which $\tilde{q}$ and $\tilde{z_a}$ 
are defined as
\bea
\tilde{q} &=& e^{-i \pi / \tilde{\t}} \nonu
\tilde{z}_a &=& - 
\left(\frac{1}{2}-  \frac{a}{N} \right) \frac{\pi}{{\tilde \t}} 
\eea
Notice that this result depends on $\tau$ only via the effective
coupling constant $\tilde{\tau}=\tau/N$. 
In the large-$N$ 't Hooft limit $\tilde{q} = e^{-\pi \l}$. 
In order to analyze the small $\l$ limit, 
we need to use the modular properties of our result in the effective
coupling $\tilde{\tau}$. In particular, performing the transformation 
$\tilde{\t} \rightarrow - \frac{1}{\tilde{\t}}$ and we find
\be
\frac{\theta_3'({\tilde z}_a| - \frac{1}{\tilde{\tau}})}{
\theta_3({\tilde z}_a| - \frac{1}{\tilde{\tau}})} =
- \tilde{\t} \frac{\theta_3'(- {\tilde \t} {\tilde z}_a |
{\tilde \tau})}{\theta_3(-{\tilde \t}{\tilde z}_a |{ \tilde \tau})}
+ \frac{2 i {\tilde \t} {\tilde z}_a }{\pi}
\ee
Then the eigenvalues can be rewritten as
\be
\l_a = 
- \frac{\theta_3'(- {\tilde \t} \tilde{z}_a|\tilde{\tau})}
{\theta_3(- {\tilde \t} \tilde{z}_a|\tilde{\tau})}
\ee
Here the anomalous piece of the $\theta$ functions under the $\tilde{S}$ 
transformation exactly cancels the leading constant term in $\l_a$.
In the small $\l$ limit $\tilde{\t}
\rightarrow  i \infty$, $e^{i \pi \tilde{\t}} \rightarrow 0$, and 
in this limit only the first term in the expansion of $\theta$ functions
survives. This corresponds to the decoupling limit where the theory
flows to the 't Hooft limit of the 
minimal ${\cal N}=2$ SUSY Yang-Mills theory with gauge
group $SU(N)$. The limiting formula for the eigenvalues is 
\bea
\l_a &=& 2 \sin (2  {\tilde \t} \tilde{z}_a) 
e^{i \pi \tilde{\tau}} \nonu &=& - 2 e^{- \pi / \l} 
\sin{\frac{2 a \pi}{N}} \label{lconf}
\eea
In this limit, the eigenvalues computed from the Lax matrix can be compared
with the results of Douglas and Shenker \cite{ds} for the minimal
${\cal N}=2$ theory. In the large-$N$ limit we can find the density
$\rho$ of the eigenvalues normalized to lie between zero
and one. The result is the Wigner distribution  
\be
\rho(\l) = \left. \frac{1}{\l'(x)} \right|_{x=x(\l)} 
= \frac{2}{ \pi \sqrt{1 -\l^2}}
\ee
This agrees with the large-$N$ limit of the distribution found by
Douglas and Shenker as computed in \cite{ferrari}
\footnote{Our normalization is such that 
$\int_{0}^1 \rho(x) d x = 1$.}.

\section{General vacua in the supergravity limit}

To compute the eigenvalue distributions in the supergravity limit, we
first examine the limit in the confining vacuum. The eigenvalues in
the confining vacua are given in (\ref{confo}), and in terms of $N$
and the 't Hooft coupling $\l$ can be explicitly written as

\be 
\l_a = \l \left[ \frac{1}{2} - \frac{a}{N}  + 
\sum_{l=1}^{\infty} 
\left( \frac{e^{- 2 \pi \l (l - \frac{a}{N} )}}{ 1 + 
e^{- 2 \pi \l (l - \frac{a}{N})}}  - 
\frac{e^{-2 \pi \l (l - 1 + \frac{a}{N} )}}{ 1 +
e^{- 2 \pi \l (l - 1 + \frac{a}{N} )}} \right) \right]
\ee

The supergravity limit is taken $N \rightarrow \infty$, while keeping
the 't Hooft coupling $\l$ fixed, large. Taking the $N \rightarrow \infty$
limit $\frac{k}{N}$ becomes a continuous parameter ranging from 0 to 1.
In this limit we have the continuos distribution

\be
\l (x) = \l \left[ \frac{1}{2} - x  +
\sum_{l=1}^{\infty}
\left( \frac{e^{- 2 \pi \l (l - x )}}{ 1 +
e^{- 2 \pi \l (l - x)}}  -
\frac{e^{-2 \pi \l (l - 1 + x )}}{ 1 +
e^{- 2 \pi \l (l - 1 + x )}} \right) \right]
\ee
When $\l$ is large the summation terms are exponentially supressed, so
in the supergravity limit the the confining vacua are
\be
\l(x) = \l \left( \frac{1}{2} - x \right)
\ee
corresponding to a constant distribution. (We note that at the endpoints
$x=0$ and $x=1$ there is a constant shift in the eigenvalues. A similar
endpoint effect can be seen when analyzing the eigenvalue distribution
from the Seiberg-Witten curve. Such a constant endpoint shift does not
effect the constant distribution of the eigenvalues.)

It is convenient to normalize the uniform distribution to range one, so in 
the following we will consider the suitably normalized eigenvalues. In
the confining vacuum this means dividing by the 't Hooft coupling $\l$,
so the distribution we arrive at the supergravity limit is
\be
\tilde{\l}(x) = \frac{1}{2} - x
\ee
where the normalized eigenvalues are denoted by $\tilde{\l}$. 

Next we examine the supergravity limit of the eigenvalues in a general
vacuum.
The set of vacua are labelled with three integers $(p, q,l)$ with
the coupling constant
\be
\tilde{\tau} = \frac{p \t + l}{q} \quad pq=N \quad l=0 \ldots q-1
\ee
The Higgs vacuum corresponds to (N, 1, 0) while the confining vacuum is
(1, N, $l$). An SL(2, Z) transformation performed on a generic vacuum
takes it into another vacuum, and also changes the coupling constant
to a different coupling $\bar{\tau}$. Then there exists another SL(2, Z)
transformation which maps this vacuum to one element of the sets of
the original vacua, characterized with different integers $(p',q',l')$.
In this way SL(2, Z) transformations map the set of vacua among themselves.

A general vacuum is th then connected to the confining vacuum by a modular
transformation with weight 1:
\be
\l  (\tau) = \frac{1}{c \t + d} \l_{conf}
\left(\frac{a \t + b}{c \t + d} \right)  \label{mod}
\ee
The weight can be seen for example from the description of vacua from
the curve, in which the Riemann zeta functions transform with modular
weight 1.

In a general vacuum the supergravity limit is taken by taking $N$ to 
infinity, while keeping the dual 't Hooft coupling fixed and large. 
For a general vaccum connected to the Higgs vacuum by a modular 
transformation (\ref{mod}) we find in the large $N$ fixed $\l_D$ limit
\be
\tilde{\l}(x) =  \frac{1}{2} - x  +
\sum_{l=1}^{\infty}
\left( \frac{e^{- 2 \pi \l_D (l - x )}}{ 1 +
e^{- 2 \pi \l_D (l - x)}}  -
\frac{e^{-2 \pi \l_D (l - 1 + x )}}{ 1 +
e^{- 2 \pi \l_D (l - 1 + x )}} \right) \label{fform}
\ee
where $\l _D$ is defined from $\tau_D = \frac{a \tau +b}{c \tau +d}$ as
$\l_D = i N / \tau_D$. Here we normalized the eigenvalues appropriately
by $\l_D / (c \tau + d)$. The modular factor comes as the weight of the
weight-1 modular transformation. In the supergravity limit the summation
is exponentially suppressed, and the eigenvalues for the general vacua are
\be
\tilde{\l}(x) =  \frac{1}{2} - x  \label{evalsugra}
\ee

In conclusion, in the supergravity limit we find a uniform distribution
for a general vacuum. The corrections to the supergravity limit are the 
summation terms with the exponential suppression in (\ref{fform}). 
These terms represent instanton contributions to the uniform supergravity
distribution coming from $(m,n)$-strings which wrap the two-sphere in
the Polchinski-Strassler geometry. 
The contribution of instantons can be rewritten by 
expanding the denominators and collecting the powers of each exponential
term $e^{-\frac{2 \pi \l}{N} m}$. In this way we arrive at the alternative
expression
\be
\tilde{\l}(x) =  \frac{1}{2} - x  +
\sum_{m=1}^{\infty} \left( \sum_{\scriptstyle d|m \atop \scriptstyle 
  d = - k \bmod N} (-1)^{\frac{m}{d} -1} 
- \sum_{\scriptstyle d|m \atop \scriptstyle d = k \bmod N}
  (-1)^{\frac{m}{d} -1} \right) 
e^{-\frac{2 \pi \l_D }{N} m} 
\ee
This form of (\ref{fform}) exhibits the coefficients the instanton 
sum explicitly.

\section{Condensates}

The physical quantities which are computed from the eigenvalue distribution
are the condensates $u_k = {\rm Tr} \Phi^k$. These essentially correspond to
the integrals of motion for the integrable system, which are expressed
as the symmetric polynomials of the eigenvalues of the Lax matrix. 

In this section we compare some of the condensates computed from 
the eigenvalues to earlier results for these. First we examine the 
semiclassical limit of the condensate $u_2 = Tr \Phi^2$ in the 
Higgs vaccum.

The precise connection between $u_2$ and the symmetric polynomial 
constructed from the squares of the eigenvalues is encoded in the 
Lax equations. In order the Lax equations to give Hamilton's equations,
the square of the function $x(u)$ in the Lax matrix is related
to the potential $V(u)$ by a constant, as described in (\ref{xlax}).
In the semiclassical limit
\bea
v_{ab} &=& V(u_{ab}) = \frac{1}{4 \sinh^2 \frac{1}{2} u_{ab}} + 
\frac{1}{12}\nonu
x_{ab} &=& X(u_{ab}) = \frac{1}{2} \coth \frac{1}{2} u_{ab}  
\eea
so in this limit
\be
C = -\frac{1}{6}
\ee
We also have
\be
I_2 = \frac{1}{2} {\rm Tr} L^2 = \sum_{a>b} x_{ab}^2 = \sum_{a>b} (v_{ab} - C)
\ee
from which
\be
u_2 = \sum_{a>b} v_{ab} = I_2 - \frac{N(N-1)}{12} = 
\frac{1}{2} \sum_{a=1}^N \l_a^2 - \frac{N(N-1)}{12}
\ee
Substituting the semiclassical result for the eigenvalues (\ref{evalsem})
\be
u_2 = -\frac{N(N^2-1)}{24} 
\ee
in perfect agreement with the semiclassical result of \cite{np}.

Next we compute the condensate $u_2$ in the confining vacuum. From
the eigenvalues in the confining limit (\ref{lconf}) we find
\be
u_2 = I_2 = \frac{1}{2} \sum_{a=1}^{N} \l_a^2  
= 2 e^{- 2 \frac{\pi}{\l}} \sum_{a=1}^{N} \sin^2{\frac{2 a \pi}{N}} 
= N e^{-\frac{2 \pi}{\l}} = N e^{\frac{2 \pi i \t}{N}}
\ee
This is in agreement with the result for $u_2$ computed in \cite{np} 
in terms of Eisenstein series
\be
U_2 = \frac{N^2}{24} \left( E_2 (\t) - \frac{1}{N} 
E_2\left(\frac{\t}{N}\right) \right)
\ee
upto a vacuum independent shift function. The relevant part of $U_2$ is
the second piece
\be
U_2' = - \frac{N}{24} E_2\left(\frac{\t}{N}\right) \approx 
N e^{\frac{2 \pi i \t}{N}}
\ee
in agreement with $u_2$ computed from the eigenvalues.

\section*{Acknowledgements}

We would like to thank Tim Hollowood and Prem Kumar for discussions.

\begin{appendix}

\section{Distribution of vacua from the Seiberg-Witten curve}

The Seiberg-Witten curve of the $\cn=2^*$ SU(N) theory can be obtained 
from IIA/M-theory brane construction \cite{witten} where one considers a brane 
system suitably constructed from a single NS5 and $N$ D4 branes. The
Coulomb branch is obtained as a branched N-fold cover of the torus 
with complex structure $\t$ 
\be
\t = \frac{4 i \pi}{g^2} + \frac{\theta}{2 \pi} = \frac{i N}{\l} +
\frac{\theta}{2 \pi}
\ee
where $\t$ is the complex gauge coupling, and $\l = g^2 N /4 \pi$ is 
the 't Hooft coupling.

The massive $\cn=1^*$ vacua are the singular points on the Coulomb
branch where the Seiberg-Witten curve maximally degenerates. At these
points, the curve becomes an unbranched cover, a torus itself, 
with complex parameter
\be
\bar{\t} = \frac{p \t + l}{q} \quad p q = N \quad l=0..q-1
\ee
A generic vacuum is thus characterized by the 3 integers $(p,q,l)$. The
Higgs vacuum corresponds to (N, 1, 0) while the confining vacuum is 
(1, N, $l$). For the confining vacua, it is enough to consider the case
$l=0$, because $l$ nonzero is connected to this by a modular transformation.

The general SW curve can be described in terms of coordinates of the 
D4-branes as
\be
\prod_{a=1}^N (v - v_a(z)) = 0
\ee
here z is on the covered torus with complex structure $\t$, which has periods 
$\t = \o_2 / \o_1$. 

The $\cn=1^*$ vacua correspond to the singular points where the curve
maximally degenerates to a torus with complex structure $\bt$. 
The coordinates at these vacua, taking the case $l=0$, are given as \cite{npt}
\bea
v_{sr}(z) &=& N m \bar{\z}(z - z_1 + 2 s \o_1 +  2 r \o_2) \nonu
&&- m \left( \sum_{t=0}^{p-1} \sum_{u=0}^{q-1} \bz(z+ 2 t \o_1 + 2 u \o_2)
+ 2 p s \bz(p \o_1) + 2 q r \bz(q \o_2) \right) 
\eea
where $s=0,1..q-1$ and $r=0,1..p-1$. $\bz$ is the Weierstrass zeta function
with complex structure $\bt$ and periods $\bar{\o}_1 = q \o_1$, and
$\bar{\o}_2 = p \o_2$, where we fix $\o_1 = i \pi$. $z_1$ denotes the 
position of the NS5 brane, and $m$ is the hypermultiplet mass.

In the following we determine the distribution of Higgs and confining
vacua in the large $N$ limit. 

In the Higgs vacuum ($N$, 1, 0) we denote the $r=0,1..N-1$ coordinates 
by $v^H_r(z)$ 
\be
\vh = N m \bar{\z}(z - z_1 + 2 r i \pi \t) - m \left( \sum_{u=0}^{N-1} 
\bz(z+ 2 u i \pi \t) + 2 r \bz(N i \pi \t) \right)
\ee
We are interested in the distribution of vacua around the their average,
so we redefine the coordinates to be centered
\be
V^h_r(z) = \vh - \bar{v}^H(z)
\ee
so that
\be
\sum_{r=0}^{N-1} V^H_r(z) =0
\ee
The new coordinates are then expressed as

\be
V^h_r(z) = N m \bar{\z}(z - z_1 + 2 r i \pi \t) - m \sum_{u=0}^{N-1} \bz(z - z_1 + 2 u i \pi \t) + m (N-1 - 2 r ) \bz(N i \pi \t) 
\ee

Taking the limit $N$ large while keeping the 't Hooft coupling $\l$ fixed, 
$\t \rightarrow i \infty$. 
The $\bz$ function
is periodic with periods
\bea
\bar{\o}_1 &=& q \o_1 = i \pi \nonu
\bar{\o}_2 &=& p \o_2 = N i \pi \t 
\eea
and its complex structure is
\be
\bt = \frac{\bar{\o}_2}{\bar{\o}_1} = N \t \rightarrow i \infty 
\ee
In this limit the $\bz$ function reduces to 
\be
\bz(z) = - \frac{z}{12} + \frac{1}{2} \coth{\frac{z}{2}}
\ee
and
\bea
V^h_r(z) &=& \frac{Nm}{2}  \coth \left( \frac{z-z_1}{2} + r i \pi \t \right) 
\nonu
&&- \frac{m}{2} \sum_{u=0}^{N-1} \coth \left(  \frac{z-z_1}{2} + u i \pi \t 
\right) + \frac{m}{2} (N-1 - 2 r) \coth \frac{N i \pi \t}{2} \label{vh}
\eea
In the $\t \rightarrow i \infty$ limit this becomes 
\bea
V^h_r(z) &=& m r - \frac{m N}{2} - \frac{m}{2} \coth \frac{z - z_1}{2}  \quad
r > 0 \nonu 
V^h_0(z) &=&  m r + \frac{m (N-1)}{2} \coth \frac{z - z_1}{2} \quad r=0
\eea
The fact that the $r=0$ coordinate is different has no significance in 
the continuous $N \rightarrow \infty$ limit, since this is of measure 
zero.

To determine a distribution for the coordinates, we scale them between 0
and 1, that is divide by $N$. Defining the parameter
\be
x= \frac{r}{N}
\ee
and 
\be
h(x) = \frac{V^h_r(z)}{N} 
\ee
we find 
\be
h(x) = m x - m/2 
\ee
Here the z-dependent piece is which is of order $1/N$ is dropped. Rescaling
the coordinates by $m$ we find the constant distribution in the Higgs vacua
in the large $N$ fixed 't Hooft coupling limit
\be
\rho_{H}(\tilde{h}) = \left. \frac{1}{\tilde{h}'(x)} \right|_{x=x(\tilde{h})}  
= 1
\label{distrH}
\ee
Here we denoted the rescaled coordinates by $\tilde{h}$.
(The mass $m$ in the integrable
system description is absorbed in the scaled variables in which the
superpotential is written. )

The constant distribution computed from the curve agrees with the constant
distribution found in terms of the eigenvalues of the Lax matrix in the
semiclassical limit (\ref{evalsem}). 

One can similarly analyze the supergravity limit of the confining (1, $N$, 0) 
vacua, with the centered coordinates
\be
V^c_s(z) 
=  N m \bz(z - z_1 + 2 s i \pi) -m \sum_{t=0}^{N-1}  \bz(z - z_1 + 2 t i \pi)
+m (N- 1 - 2 s)  \bz (N i \pi)
\ee
$\bz$ now has the periods
\bea
{\bar \o}_1 &=& N i \pi \nonu
{\bar \o}_2 &=& i \pi \t 
\eea
and complex structure
\be
\bt = \frac{{\bar \o}_2}{{\bar \o}_1} = \frac{\t}{N} \approx \frac{i}{\l}
\ee
Here the $\theta$ term in the gauge coupling is neglected. In the large $N$
limit with fixed large 't Hooft coupling $\l$ (the supergravity limit)
$\bt \rightarrow 0$. In order to get an expansion of the $\z$ function, 
a modular S-transformation has to be performed. The modular transformation
acts on $\z$ as
\bea
\t' &=& \frac{a \t + b}{c \t + d} \nonu
\z(z|\t') &=& (c \t + d) \z (z (c \t + d) | \t) 
\eea
Performing the S transformation on $\z$ when 
$\l$ is large, so in the supergravity limit, the $\z$ function 
can be replaced by its semiclassical limit. 
Then the supergravity (large $N$, large $\l$)
limit can be similarly analyzed as for the case of the Higgs vacuum.
We find a constant distribution in agreement with the distribution
computed from the integrable system (\ref{evalsugra}). Because of the 
modular properties of the $\zeta$ function, the coordinates of the 
Higgs and confining vacua as well as their large $N$ distributions are 
related by the modular transformation 
in $\tau = i N/\ \l$ as
\be
-\frac{1}{\tau} V^h_s \left(-\frac{1}{\tau}, -\frac{1}{\tau} z \right)
= V^c_s (\tau, z)
\ee

\end{appendix}

\end{document}